\documentclass[conference]{IEEEtran}
\IEEEoverridecommandlockouts
\usepackage[utf8]{inputenc}
\usepackage{cite}
\usepackage{amsmath,amssymb,amsfonts}
\usepackage{algorithmic}
\usepackage{graphicx}
\usepackage{subfigure}
\usepackage{textcomp}
\usepackage{xcolor}
\begin{document}

\title{DualSlide: Global-to-Local Sketching Interface for Slide Content and Layout Design}

\author{\IEEEauthorblockN{Jiahao Weng\IEEEauthorrefmark{1}, Xusheng Du\IEEEauthorrefmark{2}, Haoran Xie\IEEEauthorrefmark{3}}
\IEEEauthorblockA{Japan Advanced Institute of Science and Technology,
Ishikawa, Japan\\
Email: \IEEEauthorrefmark{1}s2110022@jaist.ac.jp,
\IEEEauthorrefmark{2}xushengdu@jaist.ac.jp,
\IEEEauthorrefmark{3}xie@jaist.ac.jp}}
\maketitle
\newcommand\blfootnote[1]{%
  \begingroup
  \renewcommand\thefootnote{}\footnote{#1}%
  \addtocounter{footnote}{-1}%
  \endgroup
}
\blfootnote{\IEEEauthorrefmark{3}Corresponding author.}

\maketitle

\begin{abstract}
Online learning and academic conferences have become pervasive and essential for education and professional development, especially since the onset of pandemics. Academic presentations usually require well-designed slides that are easily understood. Sketches that visually represent design intentions and are readily  accessible to the average users. To assist non-expert users in creating visually appealing academic slides, we propose DualSlide, a global and local two-stage sketching interface system that provides image retrieval and user guidance. At the global stage, DualSlide provides a heat map canvas to display the distribution of all slide layouts in a dataset, allowing users to explore the reference slides efficiently. At the local stage of the system, detailed references and guidance for designing slide content, such as diagrams and fonts, can be provided. We further propose a sketch-matching algorithm to compare the user's input sketch and similar diagrams. All user guidance can be adapted in real-time editing, and users can design slides with a high degree of freedom. We conducted a user study to verify the effectiveness and usability of the proposed DualSlide system confirming that DualSlide provides high retrieval accuracy and satisfactory design results with a good user experience.

\end{abstract}

\begin{IEEEkeywords}
two-stage design, sketching interface, slides, layout design
\end{IEEEkeywords}

\section{Introduction}
Creating visually appealing and effective slides is a challenging task, especially for users who have no experience with design. It is crucial to help users create high-quality and attractive slides which can have an impact on the audience's comprehension and retention of the information~\cite{garner2013design}. The use of consistent design elements can enhance the professionalism and clarity of slides, such as color and font selection~\cite{yi2022assessment}. However, even for experienced designers, it is challenging to create visually appealing slides because sophisticated design principles and techniques are required. Thus, there is an urgent need to assist users in the creation of well-designed documents, such as slides. 

There is a growing trend for online courses and academic conferences to require instructors and presenters to create well-designed slides for lectures and presentations. The quality of the slides can affect the audience’s understanding, but it can be time-consuming for inexperienced users to create visually appealing slides. Although presentation software, such as Microsoft PowerPoint is powerful, inexperienced users may find it more difficult to create well-designed slides than experienced designers. This is because inexperienced users are unfamiliar with design principles and may need more time to find references to inspire and inform their slide designs. Moreover, although PowerPoint provides a toolkit to generate layout suggestions for users, the toolkit may be unable to provide a consistent style and suitable suggestions for various design elements on a given slide.

\begin{figure*}[hbt]
  \centering
  \includegraphics[width=0.85\textwidth]{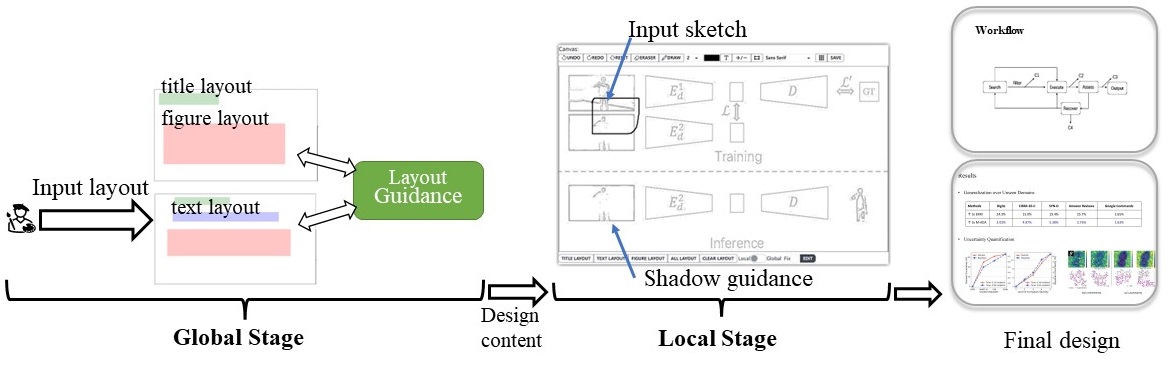}
  \caption{System concept of the proposed DualSlide sketching interface.}
  \label{fig:con}
\end{figure*}
In this work, we propose a sketching design system, DualSlide, which retrieves slide layouts and content for reference and provides users with guidance to assist them in designing slides. The proposed DualSlide system consists of two stages: a global design stage and a local design stage. This approach allows for a more structured and methodical design process, which can be beneficial to novice users who may not have experience or expertise in slide design. Dividing the task into global and local stages also helps users better understand the design process and gain a deeper understanding of the various elements that contribute to a successful design. Furthermore, this approach helps users identify and address any weaknesses in their designs. In contrast to traditional keyword-based retrieval systems \cite{weng2022hivideo}, our system can extract slides from academic presentation videos and analyze their layout using image analysis techniques. We also apply a convolutional neural network (CNN) to extract features from the slides.

The interface of the DualSlide system is divided into three parts: the drawing canvas for retrieval and editing, the heat-map canvas and shadow guidance for guidance, and the retrieval results section for displaying similar slides. Users can edit their designs on the drawing canvas as the system simultaneously retrieves similar slides for reference, and the heat-map canvas and shadow guidance provide inspiration and guidance \cite{weng2022interactive}. The retrieval results section provides a range of reference slide options from which users can choose. Overall, our system aims to make it easy for users to create visually appealing and effective slides and to help them find useful references efficiently and accurately, as shown in Fig.~\ref{fig:con}. Users first design the layout, for which the system provides references. Once the layout design is completed, users can proceed to design the details using the shadow guidance provided by the system. 

We conducted a user study to evaluate the usability of the DualSlide system. The effectiveness of the sketching interface was evaluated through a user experience experiment and a comparison experiment. In the user experience phase, a group of users were asked to create slides using the DualSlide system, and the system’s feedback and suggested slides were used as shadow guidance to further refine the design. In the comparison experiment, a group of participants were asked to create slides using either the sketching interface or a traditional slide-design method. The evaluation results verified the effectiveness of the DualSlide in contrast to traditional slide design methods.

\section{Related Work}

\subsection{Sketch Based Design Interface}
A design interface that employs abstract freehand sketches that do not contain a significant amount of visual detail has been demonstrated to be an effective way for users to express their intentions intuitively. Previous research has investigated the utilization of such sketches for a variety of tasks, including but not limited to the retrieval of web pages \cite{hashimoto2005retrieving}, the editing of images \cite{portenier2018faceshop,xiao2021sketchhairsalon}, and the generation of shadow guidance to enhance the design skills of users \cite{huang2022dualface,lee2011shadowdraw}. These techniques have also been applied to specific domains such as motion retrieval \cite{peng21}, calligraphy \cite{he20}, cartoon image generation \cite{luo21}, and facial images \cite{portenier2018faceshop}. Demonstrating their versatility and robustness. In this research, we aim to establish an interactive design interface that incorporates the use of freehand sketches to provide guidance to users in the process of designing slides.

\subsection{Layout Design and Editing}
The VINS system, as described in \cite{bunian2021vins}, employs a layout with a user interface for input in order to retrieve designs for mobile interfaces. This allows users to easily find design elements that align with their intended interface layout. However, this approach is limited to the domain of mobile interface design. Other research has applied a similar approach in the context of web design. Specifically, Hashimoto et al. \cite{hashimoto2005retrieving} proposed a method to retrieve example web pages with similar layout designs, using layout sketches as input. This approach enables users to quickly find web pages with layouts that match their intended design. However, although these previous studies have made significant contributions to the field, they are limited in scope.

\subsection{Deep Learning for Layout Analysis}
In recent years, emerging deep learning-based techniques have been proposed for PubLayNet dataset \cite{zhong2019publaynet}, which has become a widely used resource for document layout analysis. Detectron2 \cite{wu2019detectron2}, PubLayNet \cite{zhong2019publaynet}, and LayoutLM \cite{xu2020layoutlm} are some of the widespread and state-of-the-art models being used in layout analysis. These techniques have been proven effective in various research studies; notable examples include work on the analyzing and generating layouts.

\section{System Overview}
\begin{figure}[ht]
\centering 
\includegraphics[width=0.95\linewidth]{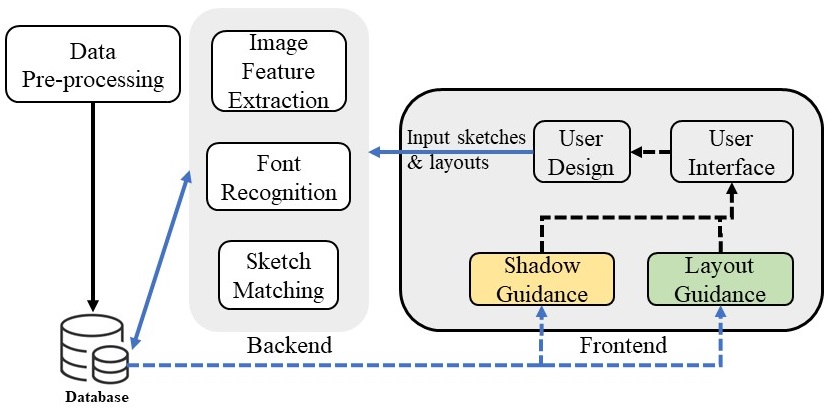}
\caption{Framework of our system. The shadow guidance is for the local design stage, and the layout guidance is for the global design stage.}
\label{fig:framework}
\end{figure}

DualSlide provides both global and local design stages for slide design. The proposed system consists of three parts: a preprocessing procedure for slide extraction; a database, which contains slide layout features and diagrams extracted from each slide, and a user interface (see Fig.~\ref{fig:framework}). To construct the database, we first extracted slides from academic conference presentation videos. We then used a layout parser to analyze and label the layout of each slide. After that, we used a convolutional neural network (CNN) to extract the features of the labeled slides and store them in the database. Additionally, we trained another neural network to recognize the font in the slides, and we improve the image similarity algorithm to implement the slide retrieval function. The user interface has three sections: the sketch canvas for user editing, the heat map canvas for design guidance, and the results section for displaying similar slides.

\subsection{Data Pre-processing}
\subsubsection{Global Design Stage}
\begin{figure}[hbt]
\centering 
\includegraphics[width=0.95\linewidth]{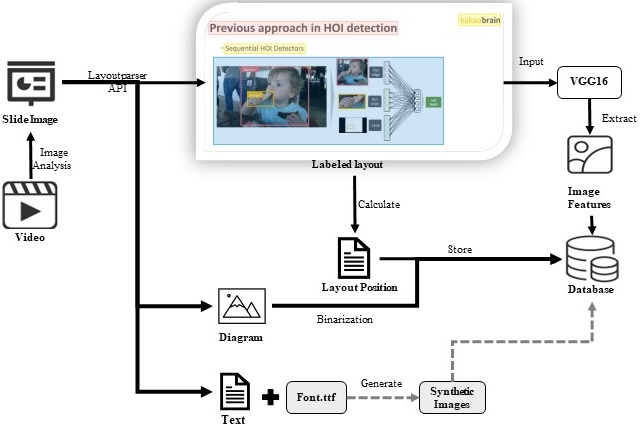}
\caption{Data pre-processing for slide extraction and slide contents extraction.}
\label{fig:dpp}
\end{figure}

In the global design stage of DualSlide, slide extraction is a crucial step as it lays the foundation for labeling layouts, calculating the distribution, and extracting slide layout features, as shown in Fig.~\ref{fig:dpp}. To extract the slide content, we compare the image hashes of each consecutive frame in an academic conference presentation video. If the difference between two frames is greater than a certain threshold, it indicates that the slide has changed, and it is therefore extracted. Once the slides are extracted, we calculate the distribution of all the slides and generate a heat map to provide a reference and inspiration to users. We then use Detectron2 \cite{wu2019detectron2} to train a model to analyze the slides, and LayoutParser \cite{shen2021layoutparser} to label the layout of each slide and extract features using a convolutional neural network. The LayoutParser tool is optimized for document image analysis tasks, allowing us to easily extract complicated document structures with just a few lines of code. After preprocessing is complete, all slide features are stored in the database and used for comparison with the user’s input sketch.

\subsubsection{Local Design Part}
In the local design stage of DualSlide, we employ LayoutParser \cite{shen2021layoutparser} to label the diagrams and texts on each slide and to extract their content. These diagrams are then binarized and stored in a database for later use in sketch matching. In addition to the diagrams, we also process the text in the slides, as shown in Fig.~\ref{fig:dpp}. We applied five of the most commonly utilized fonts in design, as outlined in previous study \cite{bernard2002comparison}. These fonts were used to generate synthetic images for later training of a font recognition model, which is used to identify the font of the text on the slide and help users design the font style.

\subsection{Sketch Matching}
Sketch matching is a crucial step in the local design stage as it lays the foundation for supporting references and generating shadow guidance. To match sketches, we propose an algorithm to compute the similarity between the input sketch and the candidate images, as shown in \eqref{eq:1}. The similarity between the images is computed by following these steps:
\begin{enumerate}
\item Initialize an empty list $M$
\item For each set of matching points $(i, j)$, compute $sim(i, j)$.
\item Use a threshold to filter the good matches. Here we use 0.75 (In order to filter out matches that may be caused by noise or outliers, a distance ratio test is applied. After extensive experimentation, we have found this to be an effective threshold for filtering out high-quality matches while minimizing the loss of important features.). If $m.distance < 0.75 \times n.distance$, append $(i, j)$ to $M$ ($m$ and $n$ are matching points, from the first image and one from the second image, respectively).
\item If $M$ is empty, return 0.
\item Otherwise, compute similarity.
\end{enumerate}

First, we assume that there are $n_1$ keypoints in the first image, each with a descriptor $d_1$, and $n_2$ keypoints in the second image, each with a descriptor $d_2$.
For each keypoint $i$ and $j$, we define their similarity as follows:
\begin{equation}
sim(i, j) = \frac{d_1(i) \cdot d_2(j)}{\left|d_1(i)\right| \cdot \left|d_2(j)\right|}\label{eq:1}
\end{equation}
where $d_1(i)$ and $d_2(j)$ represent the descriptor (extracted by the Oriented FAST and Rotated BRIEF algorithm) of the $i$-th keypoint in the first image and the $j$-th keypoint in the second image, respectively.
We use the BFMatcher algorithm (which compares each feature descriptor of one image with all feature descriptors of the other image, and returns the closest matches) to match all keypoints in the first image with all keypoints in the second image, resulting in $n_1 \times n_2$ matching points. For each matching point $(i, j)$, we select the better matching point, for example, the matching point satisfying $sim(i, j) > t$.

Finally, we calculate the similarity $S$ between the two images, defined as the average of the similarities of all better matching points:
\begin{equation}
S = \frac{\sum_{(i, j) \in M} sim(i, j)}{|M|}\label{eq:2}
\end{equation}
where $M$ is the set of better matching points: for example, the matching points satisfying $sim(i, j) > t$ (here, $t$ is a threshold value used to determine the ``good matches" among the matching points. In this case, $t$ is used in the distance ratio test, where if the distance between $m$ and $n$ is less than $0.75 \times n.distance$, then $m$ and $n$ are considered a ``good match".). 

In the above Equation \eqref{eq:2}, these matching points are referred to as ``good matches" if the matching points satisfy $m.distance < 0.75 \times n.distance$ when calculating the distance between matching points. $|M|$ represents the number of elements in the set $M$, for example, the number of good matching points. The final similarity $S$ is the average of the similarities of all ``good matching" points.

\subsection{Font Recognition}
Font recognition is used to help users select fonts for slides. When users browse reference slides, they can take the font from the slides and use it in their design. We employed a neural network to train a font recognition model. 

\subsubsection{Data Collection}
\begin{figure}[ht]
\centering 
\includegraphics[width=0.95\linewidth]{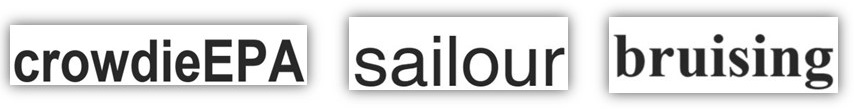}
\caption{Examples of the training data. }
\label{fig:fonts}
\end{figure}

Initially, we selected five of the most commonly utilized fonts in design, as outlined in a previously conducted study \cite{bernard2002comparison}. Since the slides in our dataset were extracted from videos, the font type of the text was not obvious. To address this, we generated synthetic data for each of the five selected fonts, consisting of 10,000 images per font (as shown in Fig.~\ref{fig:fonts}), resulting in a total of 50,000 images. Each image was labeled with its corresponding font type.

\subsubsection{Neural Network Architecture}
\begin{figure}[hbt]
\centering 
\includegraphics[width=0.95\linewidth]{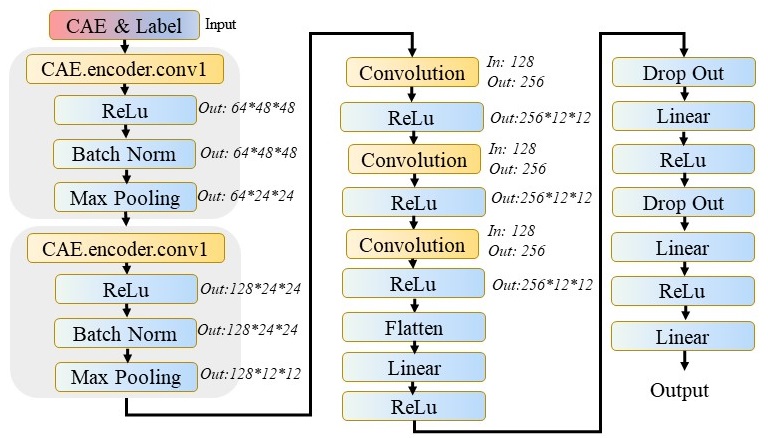}
\caption{Architecture of the font recognition model. The CNN classifier takes the output from the CAE and a number of font types as inputs.}
\label{fig:nna}
\end{figure}

The architecture of the neural network consists of a convolutional auto-encoder (CAE) \cite{cae} with a CNN classifier, as shown in Fig.~\ref{fig:nna}. The fundamental concept of a CAE is to train the encoder component of the network to extract a compressed, low-dimensional representation of the input data and then use this representation to regenerate the original input through the decoder component of the network.

\begin{figure}[ht]
\centering 
\subfigure[Architecture of CAE encoder.]{
\includegraphics[width=0.35\linewidth]{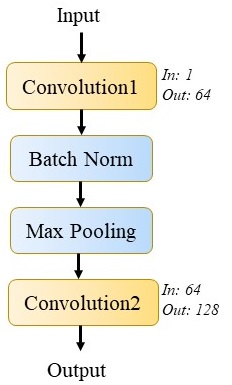}
\label{fig:caee}
}
\subfigure[Architecture of CAE decoder.]{
\includegraphics[width=0.35\linewidth]{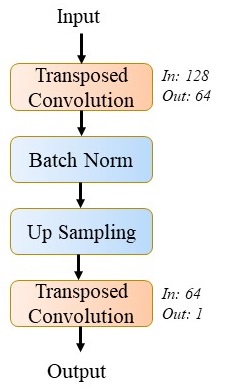}
\label{fig:caed}
}
\caption{Architecture of CAE.}
\label{fig:cae}
\end{figure}

The encoder portion of the CAE is composed of three convolutional layers. The first layer has a kernel size of 58, which reduces the dimension of the input image to 64 feature maps. The second layer is a batch normalization layer, which aims to stabilize the distribution of activations during training. The third layer is a max pooling layer with a kernel size of 2, which reduces the spatial resolution of the feature maps by a factor of 2. The final layer of the encoder is a convolutional layer with a kernel size of 3 and padding of 1, which increases the number of feature maps to 128, as shown in Fig.~\ref{fig:caee}.

The decoder portion of the CAE is composed of three transposed convolutional layers: a nearest-neighbor upsampling layer, and a batch normalization layer. The first layer is a transposed convolutional layer with a kernel size of 3 and padding of 1, which increases the spatial resolution of the feature maps to the original size. The second layer is a batch normalization layer, which aims to stabilize the distribution of activations during training. The third layer is a nearest-neighbor upsampling layer, which increases the spatial resolution of the feature maps by a factor of 2. The final layer of the decoder is a transposed convolutional layer with a kernel size of 58 and applies a sigmoid activation function, as shown in Fig.~\ref{fig:caed}.

The CNN classifier takes the output from the CAE and a number of font types as input; it has several linear and dropout layers; and it uses the output of the encoder layers as the feature map for the classifier.

The CAE architecture is trained end-to-end using a mean-squared error (MSE) loss function as the objective function, while the CNN architecture is trained using a cross-entropy loss function.

\section{Design Interface}
The proposed design interface includes both global and local design stages. The user can switch between the global and local stages by clicking the switch button shown in Fig.~\ref{fig:l-ref}.

\subsection{Global Design}\label{sec:gd}
\begin{figure}[ht]
\centering 
\includegraphics[width=0.95\linewidth]{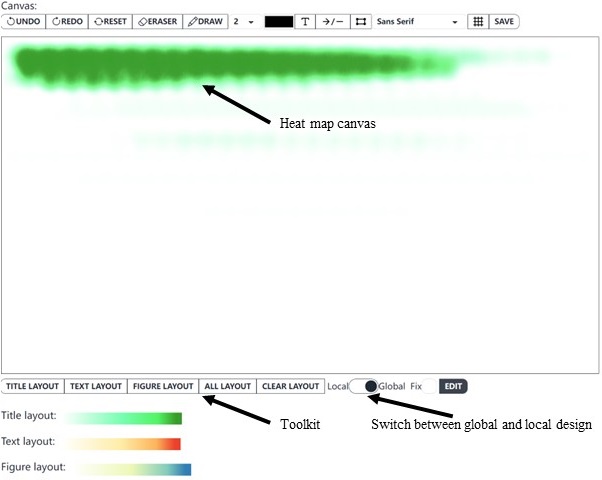}
\caption{The heat map canvas and the toolkit to change the type of heat map.}
\label{fig:l-ref}
\end{figure}

The aim of the global design is to help users efficiently and conveniently retrieve slide layouts similar to those they have designed. It also serves as a source of inspiration and guidance for slide layout design. When users want reference material for slide layout design, they can refer to the heat map, as shown in Fig.~\ref{fig:l-ref}. The features of the user's input sketches are extracted, and the Visual Geometry Group Network (VGG16) \cite{simonyan2014very} is used to compare the similarities between input features and all the features in the database. The most similar slides are displayed on the web page. The reference results change as users edit their sketches.

\begin{figure}[ht]
\centering 
\includegraphics[width=0.95\linewidth]{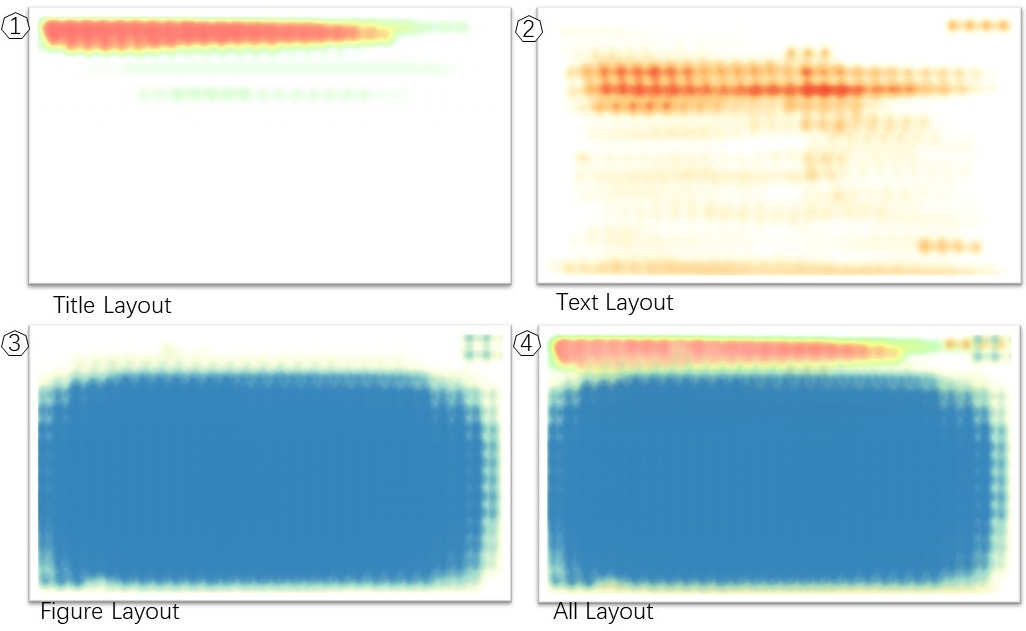}
\caption{The heat map of title layout (1), text (except title) layout (2), figure layout (3), and all of the layouts (4).}
\label{fig:hm}
\end{figure}

The heat map (shown in Fig.~\ref{fig:hm}), displays the distribution of all slide layouts in the database as a source of inspiration for users. It is divided into three sections: title, text, and figure, which cover most design scenarios. The bottom of the heat map canvas features a legend; the darker the color, the greater the distribution. Users can view the distribution of each section or the entire distribution by clicking the different toolkit buttons, as shown at the bottom of Fig.~\ref{fig:l-ref}. As the user finishes or edits their sketch, the distribution of the heat map will also change simultaneously to provide ongoing guidance on layout design.

\begin{figure}[ht]
\centering 
\includegraphics[width=0.95\linewidth]{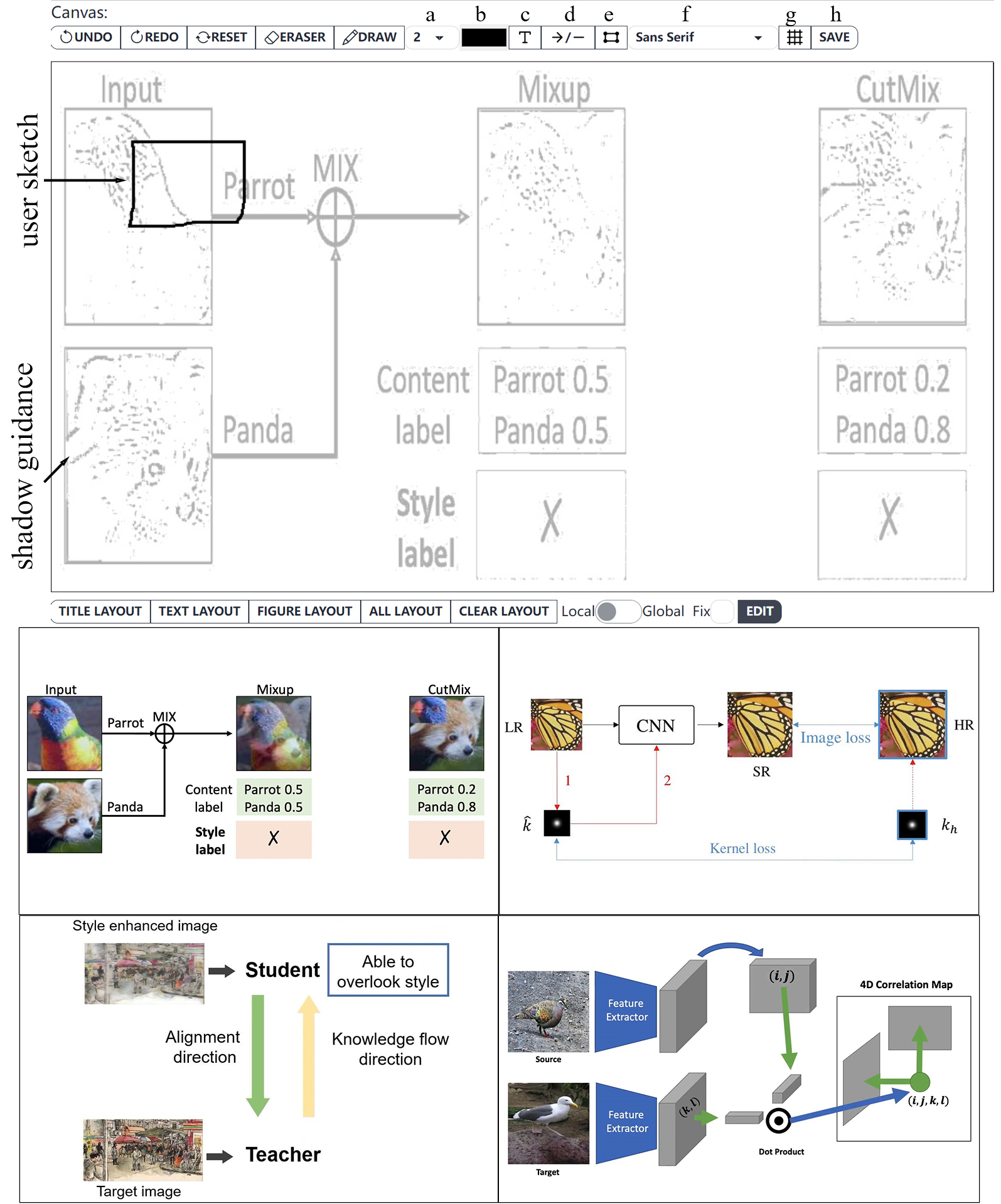}
\caption{The interface for local design. The bottom part shows the shadow guidance candidates (the visuals most similar to a user sketch). Buttons: (a) Stroke thickness, (b) Stroke color, (c) Add font, (d) Draw line or arrow, (e) Draw rectangle, (f) Choose font type, (g) Auxiliary Lines, (h) Save canvas.}
\label{fig:locui}
\end{figure}

\subsection{Local Design}
The local design is intended to help users accurately and efficiently retrieve references when they design the slide content, such as diagrams and fonts. It also serves as a source of inspiration and guidance for slide content design. For instance, when users want reference material to draw a ``framework architecture," they can first draw a rough outline. The sketch matching algorithm then compares similarities between the input sketch and the diagrams extracted from the slides. Then, the most similar diagrams are retrieved, and the first one is used as shadow guidance to provide inspiration and guidance to users, as shown in Fig.~\ref{fig:locui}. The shadow guidance also changes in real time as the user edits their sketch. Users can also click on candidate diagrams to change the shadow guidance and select a checkbox to select it. 

\begin{figure}[ht]
\centering 
\includegraphics[width=0.95\linewidth]{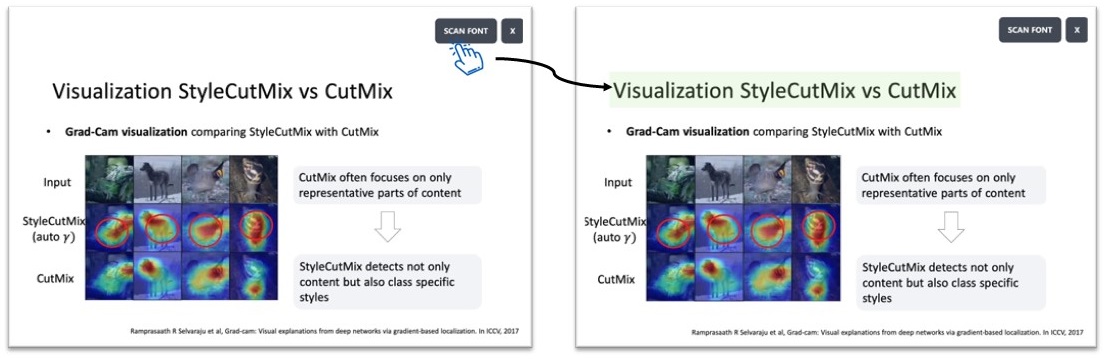}
\caption{The font recognition function. User first clicks the scan button, then labels the text they prefer, and the font style is applied to their design.}
\label{fig:font}
\end{figure}

In addition, when users click on candidate diagrams, they have another option: they can scan the text on the slide and apply the font to their design, as shown in Fig.~\ref{fig:font}.

\section{User Study}
We conducted user studies for both the global and local design stages of the proposed DualSlide system.

\subsection{Global Design}
We recruited 17 participants (college students around 20 years old, 12 males and 5 females) to participate in the experiment.

\subsubsection{Comparison Experiments}
We first compared traditional slide-design interfaces with our proposed interface. We divided 12 participants into two groups. Another five participants were recruited to evaluate the designs. The first group was asked to design three slides (focusing only on the layout) using PowerPoint (PPT), while the second group used our interface. Both groups were given an academic document (a computer science poster) to use as a reference. Once the design task was completed, the designs were evaluated by the remaining five participants based on three criteria: organization, aesthetics, and consistency.

\subsubsection{User Experience}
After the user experience experiment, we conducted a user study to assess user experiences by administering a questionnaire. The questionnaire used a 7-point Likert scale (from 1 = strongly disagree to 7 = strongly agree). We asked the 12 participants who experienced our interface in the comparison experiment to complete the questionnaire.

\subsection{Local Design}
We recruited 32 participants (12 participants to directly participate in the experiment and 20 participants to conduct the evaluation; all were college students around 20 years old, with 20 males and 12 females) to attend the experiments.

\subsubsection{Comparison Experiments}
\begin{figure}[ht]
\centering 
\subfigure[Traditional slide-design interfaces (without guidance).]{
\includegraphics[width=0.45\linewidth]{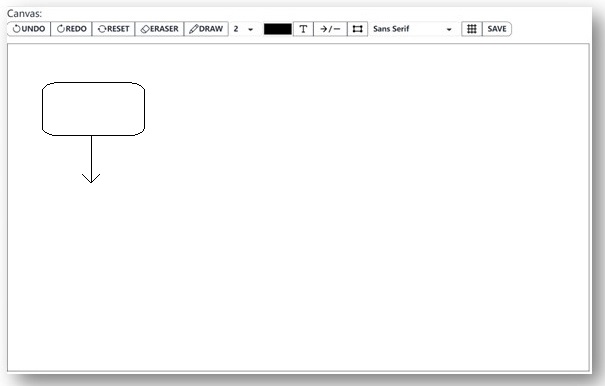}
\label{fig:ui1}
}
\subfigure[Our proposed interface (with guidance).]{
\includegraphics[width=0.45\linewidth]{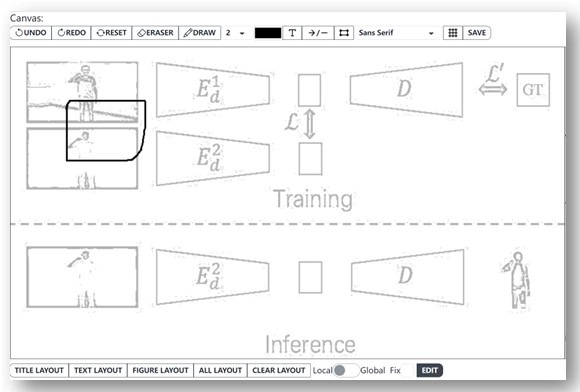}
\label{fig:ui2}
}
\caption{User interface for comparison experiment.}
\label{fig:ui12}
\end{figure}

These experiments compared traditional slide-design interfaces (Fig.~\ref{fig:ui1}) with our proposed interface (Fig.~\ref{fig:ui2}). We divided 12 participants into two groups. The other 20 participants were recruited to evaluate the designs. The first group was asked to design the framework architecture figure using UI1 (Fig.~\ref{fig:ui1}), while the second group was asked to use our interface, and vice versa. Both groups were given three selected fragments from the academic document to use as references. Once the design task was completed, the designs were evaluated by the remaining 20 participants based on three criteria: organization, aesthetics, and correctness.

\subsubsection{User Experience}
After the user experience experiment, we conducted a user study to assess user experiences by administering a questionnaire. The questionnaire used a 7-point Likert scale (from 1 = strongly disagree to 7 = strongly agree). We asked the 12 participants who experienced our interface in the comparison experiments to complete the questionnaire.

\section{Results}
\subsection{Implementation Details}
Our work was conducted on a computer system with the following hardware and software specifications. The central processing unit (CPU) was an AMD® Ryzen5 5600X CPU @ 3.70GHz X 6 with six cores and 12 threads. The graphics processing unit (GPU) was an NVIDIA RTX3060 with 12GB of video memory and support for CUDA programming. The random access memory (RAM) was 32 GB, which allowed multiple programs to run simultaneously. The operating system (OS) used was Windows 11, the latest version of Microsoft Windows. The programming language used was Python 3.8, a widely used and versatile scripting language for data analysis and machine learning. Our proposed system has an average retrieval time of 1.02 seconds for slide layout and an average computational cost of 1.33 seconds per sketch input for the sketch-matching algorithm.

\subsection{Layout Design Guidance}
\begin{figure}[ht]
\centering 
\includegraphics[width=0.95\linewidth]{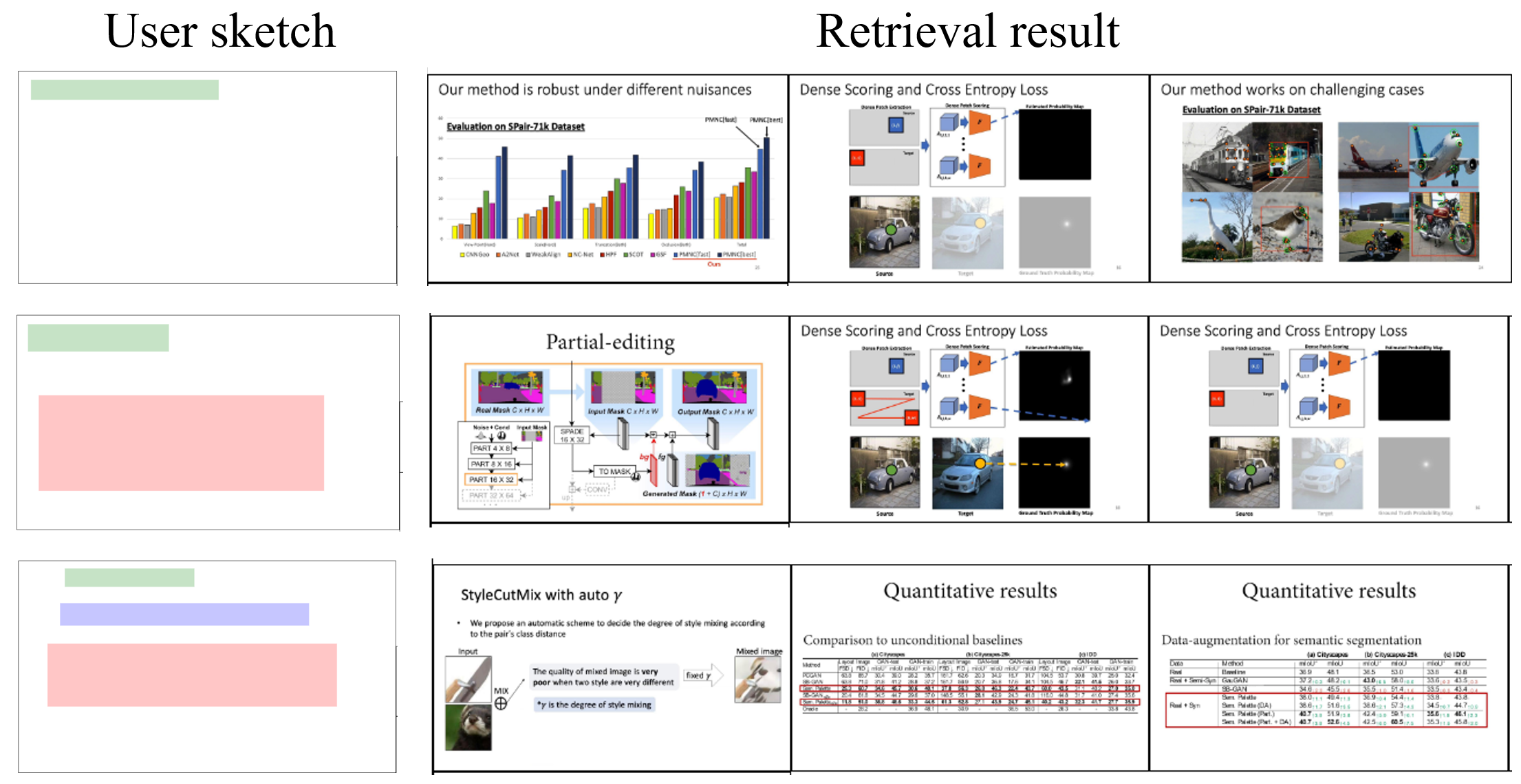}
\caption{Examples of the designed results by participants during the user study. The red part is the diagram layout, the blue part is the text layout, and the green part is the title layout. }
\label{fig:layoutG}
\end{figure}

For the local design stage, we conducted an experiment to examine how well our system retrieves similar slides based on user input sketches. As shown in Fig.~\ref{fig:layoutG}, different input sketches can retrieve different slides. Even if users do not complete the entire layout, the system will retrieve the most similar slides based on the sketches provided. The top part of Fig.~\ref{fig:layoutG} shows that even when the user only sketches the title layout, the system still performs well.

\subsection{Font Recognition}
In our research, 75 $\%$ of the data was used for training the model, while the remaining 25 $\%$ was used for validation. This split allows for the model to be trained on a large portion of the data while also being able to evaluate its performance on unseen data.

\begin{figure}[htb]
\centering 
\includegraphics[width=0.8\linewidth]{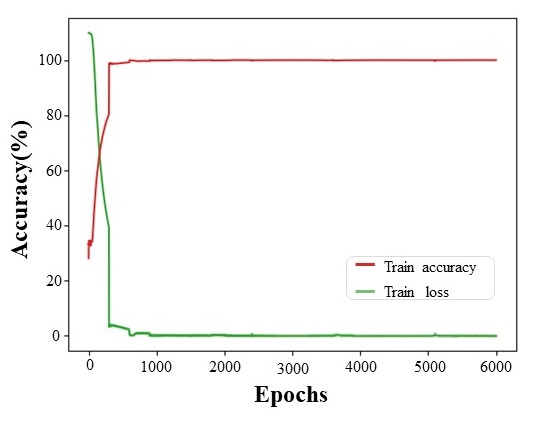}
\caption{Training results of font recognition model. The y-axis is the accuracy percentage, the x-axis is the training loss.}
\label{fig:mr}
\end{figure}

The results for our model are shown in Fig.~\ref{fig:mr}, where it can be seen that the model performed well on the training set, achieving high accuracy. This indicates that the model can perform well on font classification tasks.

\subsection{User Study}

\subsubsection{Comparison Experiment}
\begin{figure}[ht]
\centering 
\subfigure[Global design stage.]{
\includegraphics[width=0.8\linewidth]{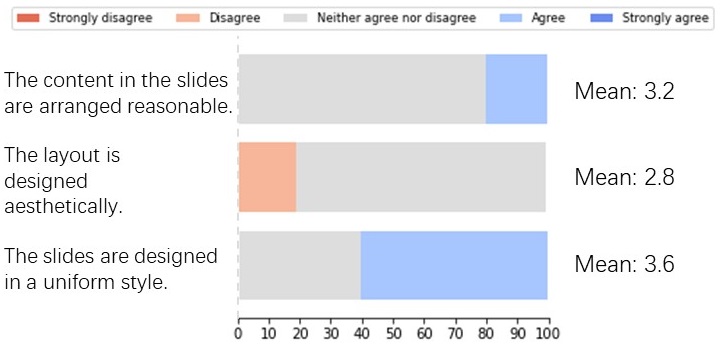}
\label{fig:layoutR}
}
\subfigure[Local design stage.]{
\includegraphics[width=0.8\linewidth]{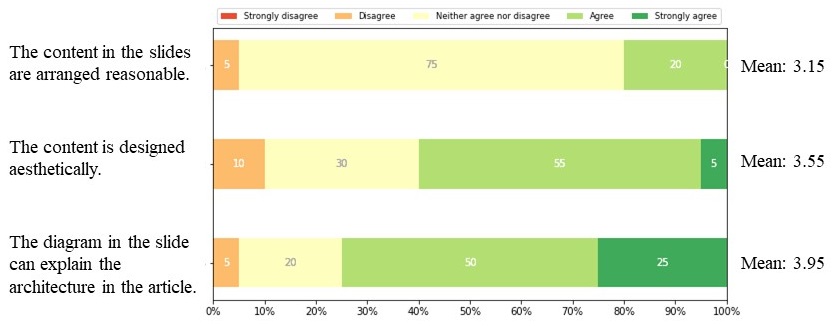}
\label{fig:contentR}
}
\caption{Evaluation results for the comparison experiments.}
\end{figure}

To quantify the ability to support users in designing slide layouts with more consistent styles, we conducted a comparison experiment. The results of this experiment are presented in Fig.~\ref{fig:layoutR}. The results demonstrate that the proposed interface is capable of helping users design slides with a high degree of consistency when compared to traditional interfaces. This is an important finding, as it suggests that the proposed interface can effectively guide users to create visually coherent slide designs, which can enhance the overall effectiveness of the resulting design.

Additionally, the results of the user comparison experiment, as shown in Fig.~\ref{fig:contentR}, indicate that the proposed interface has a positive impact on users' ability to effectively convert complex information into an aesthetically pleasing diagram. This is achieved by providing users with shadow guidance. The results of the experiment demonstrate that the proposed interface is effective in improving the aesthetics and correctness of slide content compared to traditional interfaces.

\begin{figure}[ht]
\centering 
\subfigure[User study of global design stage.]{
\includegraphics[width=0.95\linewidth]{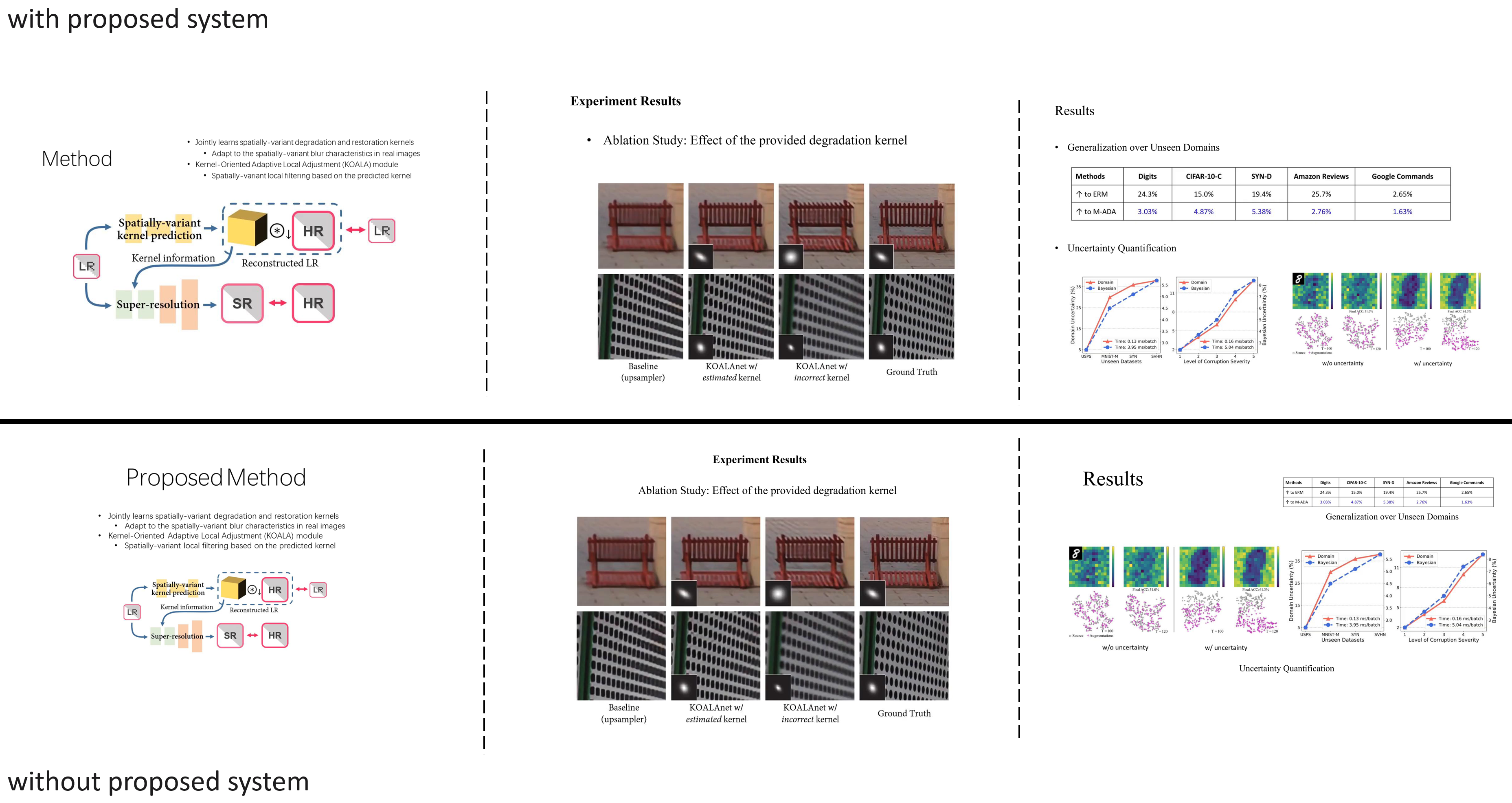}
\label{fig:layoutRbU}
}
\subfigure[User study of local design stage.]{
\includegraphics[width=0.95\linewidth]{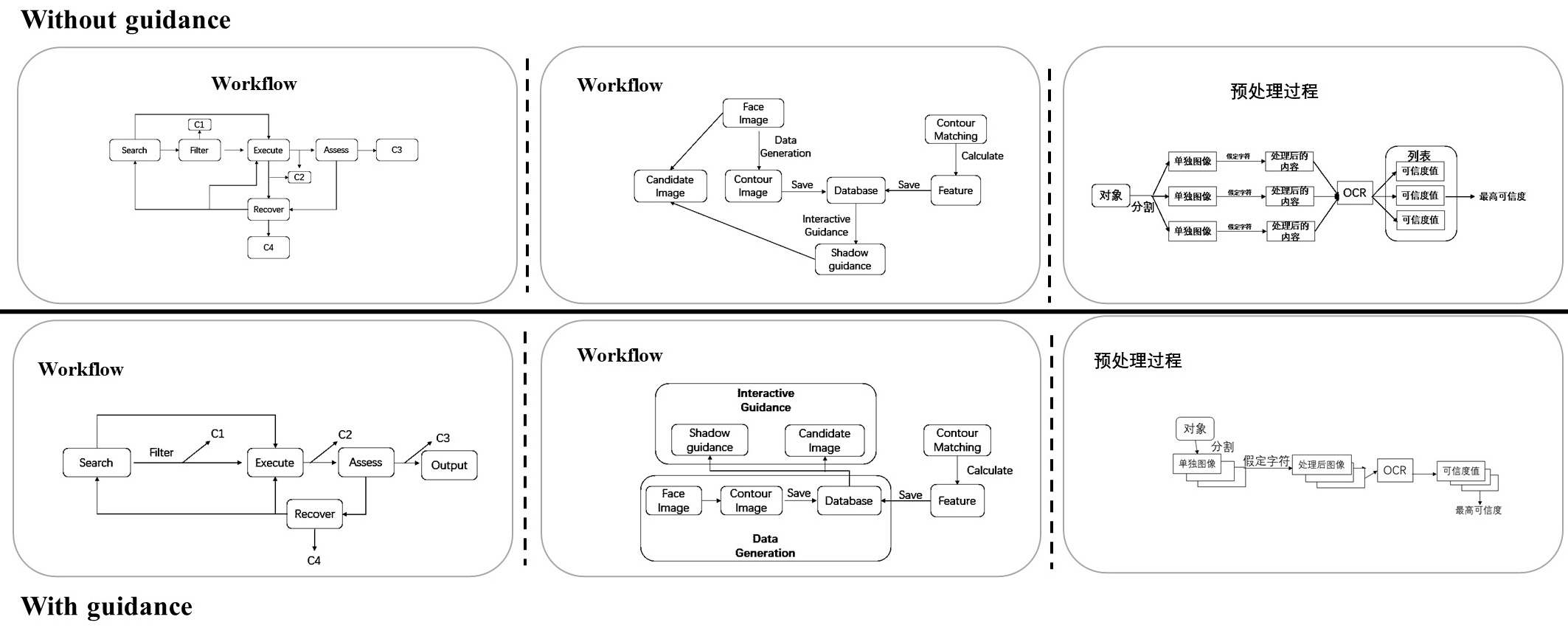}
\label{fig:contentRbU}
}
\caption{Examples of the results designed by participants.}
\end{figure}

To further illustrate the usefulness of our interface, Fig.~\ref{fig:layoutRbU} displays several examples of the slides generated by participants during the experiment for the global design stage. In this stage, we provide the participant with images and a fixed font style. The participant only needs to focus on the design layout. Fig.~\ref{fig:contentRbU} displays several examples of the slides generated by participants during the experiment for the local design stage, in which the participants need to design the diagram and font style by themselves.

\subsubsection{User Experience}
\begin{figure}[ht]
\centering 
\subfigure[Results of the questionnaire (global design stage).]{
\includegraphics[width=0.95\linewidth]{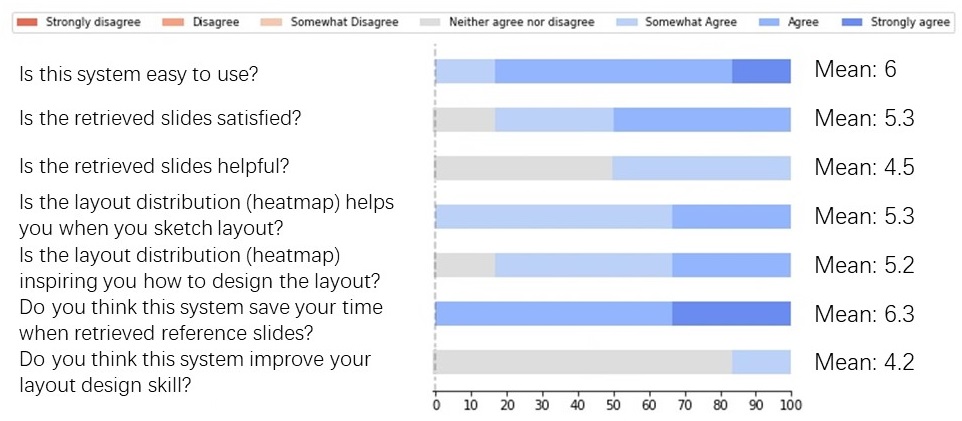}
\label{fig:layoutRUE}
}
\subfigure[Results of the questionnaire (local design stage).]{
\includegraphics[width=0.95\linewidth]{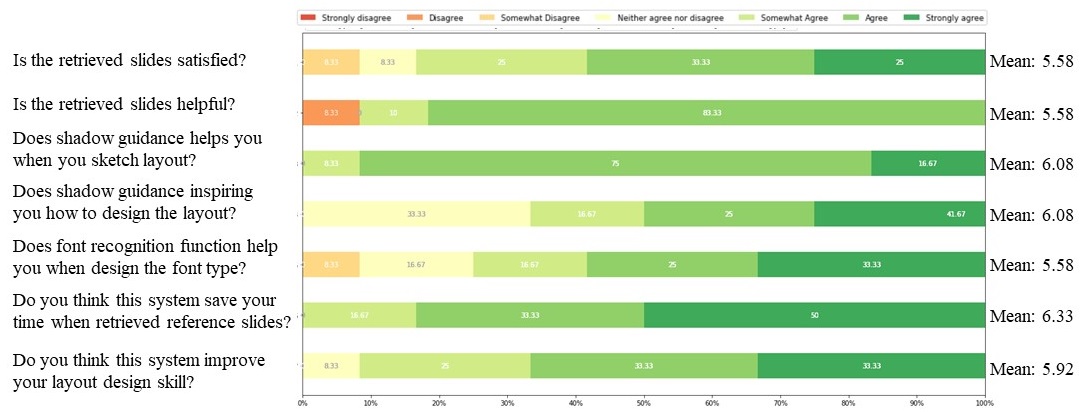}
\label{fig:contentRUE}
}
\caption{Results of the questionnaire.}
\end{figure}

To evaluate the effectiveness of the proposed interface, we conducted a user experience experiment. The results of the experiment for the global stage, as shown in Fig.~\ref{fig:layoutRUE}, indicates that not only were participants satisfied with the slides that were designed using our interface, but they also reported that the interface saved them time during the retrieval process. Additionally, the heat-map design feature, which was implemented as a means to guide users, was found to be particularly inspiring for participants in terms of layout design.

The results of the experiment about the local stage, as shown in Fig.~\ref{fig:contentRUE}, indicate that not only were participants satisfied with the slides that were designed using our interface, but they also reported that the interface saved them time during the retrieval process. Additionally, the shadow guidance and font recognition features, which were implemented to guide users, were found to be particularly inspiring for participants in terms of local design. Moreover, most participants believed that our interface succeeded in improving their design skills.

\section{Conclusion}
In this work, we proposed an interactive design system, DualSlide, that uses a heat map canvas and shadow guidance to provide users with references and guidance during slide design. A font recognition model was also included to ensure consistent font styles across slides. A user study was conducted to compare the proposed interface with traditional interfaces, and the results show that the proposed interface is effective in improving the overall design process for slide design. The results indicate that DualSlide has the potential to be generalized to a wide range of use cases, making it a valuable tool in the field of document layout and content design.

This work has some limitations, in particular, the limited size of the dataset, which results in a relatively long retrieval time of approximately 1.08 seconds per search. To address this issue, we intend to expand the dataset and improve the sketch-matching algorithm to reduce the retrieval time. We also plan to extend our research to other document types, such as posters and papers. In the current implementation of the proposed DualSlide framework, aesthetic judgment of the collected dataset and design relied on the user's discretion. To address this issue, we plan to add an assessment module that can automatically score the references provided by the system and the designed results created by users in the future.

\section*{Acknowledgment}

This research was supported by the JAIST Research Fund and JSPS KAKENHI JP20K19845, Japan.

\bibliographystyle{IEEEtranS}
\bibliography{ref}

\end{document}